\documentclass[12pt]{article}
\begin{document}
\begin{center}
{\large {\bf What is the present-day status of \\ The Copernican Principle?}}\\[4mm]
{\large {\bf Y. Nutku}}\\
Feza G\"ursey Institute P.O.Box 6 \c{C}engelk\"oy, Istanbul 81220 Turkey\\
email: nutku@gursey.gov.tr\\ December 17, 2004\\
for the Gravity Essay Competition 2005\\[2mm]
\end{center}

\begin{center}
{\large {\bf Summary}}
\end{center}
I point out that according to the Copernican principle our
universe is not unique. The way to make sense out of this
statement is for us to construct a gravitational instanton that
will tunnel out of our vacuum into another, to form a universe
other than our Hubble bubble.

\begin{center}
{\large {\bf Introduction}}
\end{center}

\noindent The Copernican Principle plays the role of a guiding
principle in all fundamental theories of physics and in our lives.
Arthur K\"ostler's chapter on Copernicus in {\it Sleepwalkers} is
entitled ``The Timid Cannon" owing to the fact that he felt the
revolutionary idea he espoused was not acceptable to the public,
let alone the authorities that be.

Quite apart from its original formulation that the earth is not
the center of the solar system, the Copernican principle has
reappeared under different guises throughout the history of
science. Its main message that we are not privileged has led to
{\it principles of denial} that must be respected in formulating
fundamental theories of physics. What we came to call principles
of denial are in fact basic truths which we do not commonly teach
our children as we bring them up. Perhaps the most striking one
among them is in thermodynamics where, roughly put, we start with
the fact that ``there is no free lunch."

Newton's formulation of the laws of mechanics as well as
Maxwell-Yang-Mills's field theories of electrodynamics and
Einstein's theory of gravity are based on principles of
gauge-invariance and general covariance which are principles of
denial that we have inherited from Copernicus. To every principle
of denial there corresponds a group of invariance and the field
equations of the physical theory must be formulated to transform
covariantly under this group.

\begin{center}
{\bf Our universe, itself, is not unique}
\end{center}

The Copernican principles of denial have been so successful in
constructing fundamental theories of physics that we have come to
regard its original formulation as a historical artifact that has
little relevance to the physics of today. Yet, as I shall
presently point out, it is still relevant in the context of its
original subject, namely cosmology. Soon after Einstein published
the general theory of relativity there appeared two of its exact
solutions due to Schwarzschild and Friedman that surprised him by
their simplicity. Both of these solutions are crucially important
for problems of cosmology even though the relevance of the former
was recognized only relatively recently. Nowadays a great deal of
effort is spent on discussing whether or not ours is a Friedman
universe that has closed, open, or flat space sections. Basically
this is a question that is only of experimental interest.

I would like to point out that for cosmology the important
observation is that our universe, itself, is {\bf not} unique
which surely is the present-day formulation of the Copernican
principle.

At first this appears to be an empty statement since we cannot be
in communication with any observers in other universes. However,
we can make sense of it by designing an experiment, albeit a {\it
gedanken} one for the present, that will enable us to create an
{\it instanton} to tunnel out of our vacuum into another and
subsequently form a new universe.

Tunnelling between two different vacua is an idea that first
appeared in Yang-Mills theories. The ``particles" mediating this
transition are localized in time, rather than the familiar ones
localized in space. Thus they are bound states of the theory with
Euclidean signature and are called instantons. There are many
parallels between gauge theories and general relativity, so it is
tempting to consider gravitational instantons which are solutions
of the vacuum Einstein field equations with Euclidean signature.
However, we cannot do so naively.

It was shown by Utiyama that Riemannian geometry of space-time can
be understood as a gauge theory for the Lorentz group along the
lines Yang and Mills had put forth. However, from the perspective
of gauge theory the field equations of Maxwell and Yang-Mills are
unlike those of Einstein owing to a coincidence between the
dimension of the Lorentz group and the space of $2$-forms on
space-time. This leads to the possibility of constructing a new
Lagrangian, Hilbert's Lagrangian for gravity which is radically
different from the Lagrangian for gauge theories. It is an
important point to keep in mind when we are considering
gravitational instantons as opposed to Yang-Mills instantons. In
the Yang-Mills case the transition between different vacua is
constructed using analytic continuation. In contrast, for the case
of gravity Calabi has shown that Ricci-flat Riemannian spaces are
governed by the elliptic complex Monge-Amp`ere equation. It is
remarkable that for Euclidean signature there is a single equation
that replaces all of the Einstein field equations. Furthermore
Chern, Levine and Nirenberg had shown that the parabolic complex
Monge-Amp`ere equation replaces Laplace's equation in the theory
of functions of many complex variables. The upshot of this is that
solutions of the complex Monge-Amp`ere equation are notoriously
not amendable to analytic continuation. Thus, except in a few very
special cases, starting with exact solutions of Euclidean
signature the passage to Lorentzian signature remains as a
fundamental difficulty. We must radically refine our ideas about
tunnelling between different vacua in the case of gravity.

The crucial gravitational instanton is the $K3$ surface that
Kummer introduced over a century and half ago. Even though we know
a great deal of its properties from the powerful index theorem of
Atiyah and Singer, we were no closer to knowing its explicit form
than Kummer's characterization of it as a quartic in complex
projective space of $3$ dimensions. This situation can be compared
to a hypothetical historical scenario for black holes. If we had
not known the Kerr solution, we would still be able to prove
theorems about black holes. But without knowledge of the exact
geometry of a black hole, it would be very difficult to know its
precise nature.

The principal difficulty in the construction of the metric on $K3$
lies in the fact that it admits no continuous symmetries. In view
of the importance of $K3$ as the most important gravitational
instanton which will radically Copernicize our ideas about
cosmology, my friends Andrei Malykh, Misha Sheftel' and I embarked
on a long project to study {\it non}-invariant solutions of the
complex Monge-Amp`ere equation. Recently this has borne fruit
\cite{mns1}-\cite{mns3} and we were able to construct
anti-self-dual metrics, which are therefore Ricci-flat, without
any Killing vectors. The method of constructing the solution, the
use of partner symmetries, is a novel one in mathematical physics.
At about the same time Maciej Dunajski and Lionel Mason \cite{dm},
also close friends from Penrose's group, had introduced ideas
which in spirit are very similar to our method.

At the moment it is not clear whether, or not some parts of our
solutions characterize the metric on $K3$. However, we now have a
novel method to construct non-invariant solutions of non-linear
partial differential equations. The issue of the relevance of
these solutions to $K3$ which will play such an important role in
our present-day idea of the Copernican principle is still under
active investigation.

\end{document}